\documentclass[journal=nalefd,manuscript=article,layout=twocolumn]{achemso}

\newcommand{\bdi}{\mathbf{i}}
\newcommand{\bdj}{\mathbf{j}}

\title{Proposal for Manipulation of Majorana Fermions in Nano-Patterned Semiconductor-Superconductor Heterostructure}
\author{Long-Hua Wu}
\affiliation{International Center for Materials Nanoarchitectonics (WPI-MANA), National Institute for Materials Science,
Tsukuba 305-0044, Japan}
\alsoaffiliation{Graduate School of Pure and Applied Sciences, University of Tsukuba, Tsukuba 305-8571, Japan}
\author{Qi-Feng Liang}
\author{Zhi Wang}
\affiliation{International Center for Materials Nanoarchitectonics (WPI-MANA), National Institute for Materials Science,
Tsukuba 305-0044, Japan}
\author{Xiao Hu}
\email{Hu.Xiao@nims.go.jp}
\affiliation{International Center for Materials Nanoarchitectonics (WPI-MANA), National Institute for Materials Science,
Tsukuba 305-0044, Japan}
\alsoaffiliation{Graduate School of Pure and Applied Sciences, University of Tsukuba, Tsukuba 305-8571, Japan}
\date{\today}

\begin{document}
\begin{abstract}
  We investigate a heterostructure system with a spin-orbit coupled semiconductor sandwiched by an $s$-wave
  superconductor and a ferromagnetic insulator, which supports Majorana fermions (MFs) at the superconducting vortex
  cores. We propose a scheme of transporting and braiding the MFs, which only requires application of point-like gate
  voltages in a system with nano-meter patterns. By solving the time-dependent Bogoliubov-de Gennes equation
  numerically, we monitor the time evolutions of MF wave-functions and show that the braiding of MFs with non-Abelian
  statistics can be achieved by adiabatic switching within several nano seconds.
\end{abstract}

\textbf{Introduction.}
  Majorana fermions (MFs) are particles equivalent to their own anti-particles \cite{bib:em}. Great research effort
  has been devoted to searching for MFs in condensed matter systems in the last decade since the particles can be
  used for constructing topological quantum computers \cite{bib:kitaev2,bib:cn_rmp,bib:wilczek}.
  Systems that attract most attentions include Pfaffian $\nu = 5/2$ fractional quantum Hall state \cite{bib:moore},
  chiral \textit{p}-wave superconductors (SCs) \cite{bib:read}, one-dimensional (1D) spinless SCs \cite{bib:kitaev1},
  superfluidity of cold atoms \cite{bib:tewari,bib:sato}, to name a few. Recently, two heterostructure systems are
  suggested as possible MF hosts, namely the topological insulator in proximity to $s$-wave SC
  \cite{bib:fu_ti} and the spin-orbit coupled semiconductor (SM) sandwiched by \textit{s}-wave SC and
  ferromagnetic insulator (FI) (SC/SM/FI) \cite{bib:sau_dev,bib:alicea,bib:linder}. At the same time, a 1D semiconductor
  nanowire with spin-orbit coupling in proximity to $s$-wave SC was investigated both theoretically and
  experimentally for realizing MFs \cite{bib:lutchyn,bib:oreg,bib:hassler,bib:potter,bib:sau_nano,bib:heck,bib:mourik}.
  Lately, a promising signal of MFs has been captured in the device of InSb nanowire/$s$-wave SC
  \cite{bib:mourik}. In these hybrid systems, the \textit{p}-wave pairing are superseded by the interplay between
  proximity-induced $s$-wave superconductivity and strong spin-orbit coupling.

  \begin{figure}[ht]
    \centering
    \includegraphics[width=\linewidth]{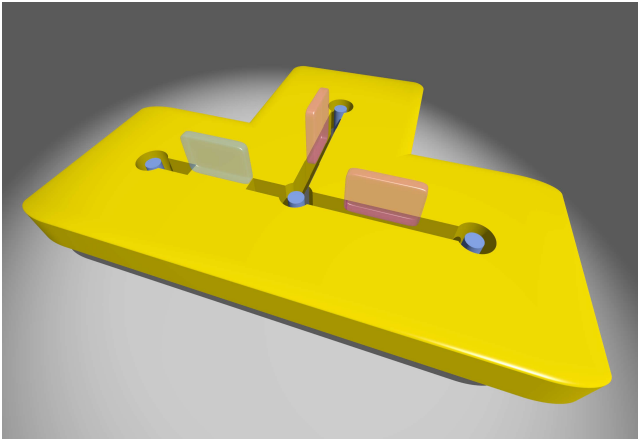}
    \caption{Schematic device setup for braiding MFs at vortex cores. There are four holes in SM layer (yellow platform)
      with one superconducting vortex (blue cylinder) pinned right beneath each of them. The electrodes
      at high-voltage states (pink rectangular prisms) prohibit electron hoppings in the regions below them, and thus
      connect effectively the holes; the blue rectangular prism denotes an electrode at zero-voltage state.}
    \label{fig:setup}
  \end{figure}

  It was revealed that MFs appear inside vortex cores in a chiral \textit{p}-wave SC \cite{bib:read}, and
  that non-Abelian statistics can be achieved by exchanging positions of vortices hosting MFs
  \cite{bib:ivanov}. However, it is difficult to manipulate vortices in experiments, which may hinder the
  realization of this genius idea. To circumvent this problem, MFs at sample edges of topological SC have been
  considered \cite{bib:liang}. Making use of the topological property, edge MFs can be braided with the desired
  non-Abelian statistics merely by tuning point-like gate voltages on links among topological SC samples.
  In order to make the edge MFs stable, one needs to embed the
  device into a good insulator. The size of topological SCs should also be chosen carefully since the wave-functions of
  edge MFs become too dilute for large samples, which makes edge MFs fragile due to excited states.

  In this work we demonstrate that the core MFs can be liberated from vortices, transported and braided by applications
  of local gate voltages. This scheme takes full advantages of SC/SM/FI heterostructure in the way shown schematically
  in Figure \ref{fig:setup}: four holes are punched in the SM layer and one vortex is induced and pinned right beneath
  each hole in the common superconductor substrate; three electrodes are placed above the small regions between holes,
  and the ones at high voltage state (pink rectangular prisms in Figure \ref{fig:setup}) connect holes by killing
  electron hoppings in SM locally. The linear dimension of holes and separation between holes should be in the regime
  of tens of nano-meters \cite{bib:liang}: for too large holes, the energy gap between the zero-energy MFs and the
  lowest excited states becomes very small, which limits the operation temperature, while for too small holes especially
  at the part of keyhole tail MF wave-functions interact with each other, which destroys the MF ground state.

  The key observation is that the geometric topology of the SM layer can be controlled by local gate voltages, and that
  when even number of holes are connected, core MFs fuse into quasi-particle states with finite energies, while one core
  MF exists when odd number of holes are connected, since the unified perimeter includes even and odd number of vortices
  in the two cases respectively.  Core MFs can then be liberated from and transported among vortices with a sequence of
  turning on and off gate voltages at the electrodes. In order to keep the topological protection, one needs to switch
  gate voltages in an adiabatic way. To monitor the dynamics of MFs, we solve numerically the time-dependent
  Bogoliubov-de Gennes (TDBdG) equation with high precision. It is then found that, with reasonable materials
  parameters, adiabatic manipulations of MFs and thus non-Abelian statistics can be achieved within several nano
  seconds, which reveals that the present scheme has the application potential for topological qubits and topological
  quantum computation.

\textbf{Topological superconductivity.}
  We start from the tight-binding Hamiltonian of a SM with Rashba-type spin-orbit coupling on a square
  lattice in proximity to a ferromagnetic insulator \cite{bib:liang}
  \begin{eqnarray}
    &&\hspace{-.9cm}H_0 = -\sum_{\bdi,\bdj,\sigma} t_{\bdi\bdj}\hat{c}_{\bdi\sigma}^\dagger \hat{c}_{\bdj\sigma} -
    \mu\sum_{\bdi,\sigma}\hat{c}_{\bdi\sigma}^\dagger \hat{c}_{\bdi\sigma} +
    \sum_{\bdi}V_z\left(\hat{c}_{\bdi\uparrow}^\dagger \hat{c}_{\bdi\uparrow}-\hat{c}_{\bdi\downarrow}^\dagger
    \hat{c}_{\bdi\downarrow}\right) \nonumber  \\
    &&\hspace{-1.1cm}+\sum_{\bdi,\sigma,\sigma'}\left\{ i
      t_\alpha^\bdi\left[\hat{c}_{(\bdi+{\bf \hat{x}})\sigma}^\dagger \hat{s}_y^{\sigma\sigma'}\hat{c}_{\bdi\sigma'} -
    \hat{c}_{(\bdi+{\bf \hat{y}})\sigma}^\dagger \hat{s}_x^{\sigma\sigma'} \hat{c}_{\bdi\sigma'}\right] + h.c.\right\},
    \label{eq:tb}
  \end{eqnarray}
  where $t_{\bdi\bdj}$ and $t_\alpha^\bdi$ are the nearest-neighbor hopping rates of electrons with reserved and flipped
  spin directions; $\mu$ and $V_z$ are chemical potential and strength of Zeeman field respectively;
  $c_{\bdi\sigma}^\dagger$ creates one electron with spin $\sigma$ at lattice site $\bdi$; $\vec{s} =
  (\hat{s}_x,\hat{s}_y,\hat{s}_z)$ are the Pauli matrices.

  The proximity-induced superconductivity in SM is described by
  \begin{equation}
    H_{\mathrm{sc}} = \sum_{\bdi}\left(\Delta_\bdi\hat{c}_{\bdi\uparrow}^\dagger\hat{c}_{\bdi\downarrow}^\dagger + h.c.\right),
  \end{equation}
  where $\Delta_\bdi$ is the pairing potential at site $\bdi$.
  The BdG equation is then given by
  \begin{equation}
    \left(
      \begin{array}{cc}
      H_0 & \Delta \\
      \Delta^\dagger & -\hat{\sigma}_y H_0^*\hat{\sigma}_y \\
    \end{array} \right)
    \Psi(\vec{r}) = E\Psi(\vec{r})
    \label{eq:BdG}
  \end{equation}
  with the Nambu spinor notation \(\Psi(\vec{r}) = \left[u_\uparrow(\vec{r}), u_\downarrow(\vec{r}),
    v_\downarrow(\vec{r}), -v_\uparrow(\vec{r})\right]^{T}\), which defines the quasi-particle operator
    \(
      \gamma^\dagger  = \int d\vec{r} [\sum_\sigma u_\sigma(\vec{r})\hat{c}_\sigma^\dagger(\vec{r}) +
      v_\sigma(\vec{r})\hat{c}_\sigma(\vec{r})].
    \)

  \begin{figure}[ht]
    \centering
    \includegraphics[width=\linewidth]{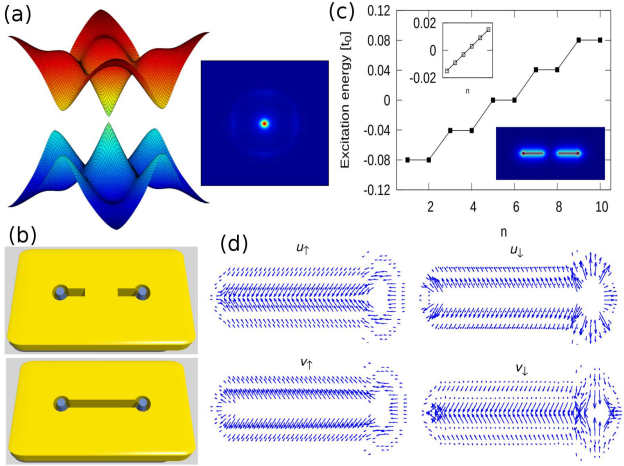}
    \caption{(a) Dispersions of two bands close to Fermi level (left panel) and Berry curvature $\nabla\times{\cal
      A}$ (right panel) in the first Brillouin zone.
      (b) System of two holes and two vortices with holes isolated (top panel) and connected (bottom panel).
      (c) Energy spectrum of several low-energy excitations at vortex cores when the holes are isolated. Upper inset: energy
      spectrum of excitations at the sample edge. Lower inset: distributions of zero-energy quasi-particles. Results are for
      $\Delta=0.5t_0$, $V_z=0.8t_0$, $\mu=-4t_0$ and $t_\alpha=0.9t_0$ with a sample of $400\times 200$ sites.
      (d) Four spinor components of zero-energy states at the right hole, with the length and azimuth angle of vectors
      denoting the amplitude and phase of spinor fields respectively.}
      \label{fig:c2}
  \end{figure}

  To reveal the topological nature of the system, we evaluate the Chern number (also known as the TKNN index)
  \cite{bib:TKNN}. For this
  purpose, we treat first the system in ${\bf k}$-space on the basis $\left[c_{{\bf k}\uparrow}^\dagger, c_{{\bf
  k}\downarrow}^\dagger, c_{-{\bf k}\uparrow}, c_{-{\bf k}\downarrow}\right]^T$. The Hamiltonian then reads
  \begin{equation}
    \hspace{-.8cm}
    H_{\bf k} =\left(
    \begin{array}{cc}
      \varepsilon({\bf k})+V_z\hat{s}_z + \vec{R}({\bf k})\cdot\vec{s} & i\Delta({\bf k})\hat{s}_y \\
      -i\Delta^*({\bf k})\hat{s}_y & -\varepsilon({\bf k})-V_z\hat{s}_z +\vec{R}({\bf k})\cdot \vec{s}^* \\
    \end{array}\right),
  \end{equation}
  where $\varepsilon({\bf k}) = -2t_0(\cos k_x + \cos k_y)-\mu$, $\vec{R}({\bf k}) = (-2t_\alpha \sin k_y, 2t_\alpha \sin
  k_x)$.
  Diagonalizing $H({\bf k})$, we obtain the eigenstates $\Phi(\bf k)$ and band structure $E({\bf k})$ in the first
  Brillouin zone $T^2  = [-\pi/a,\pi/a) \otimes [-\pi/a,\pi/a)$,
   as shown in Figure \ref{fig:c2}(a). The band topology is characterized by the Chern number
  \begin{equation}
    c = \frac{1}{2\pi i}\int_{T^2} d^2{\bf k} \cdot\left(\nabla \times {\cal A}\right)
    \label{eq:chern}
  \end{equation}
  with Berry connection for all occupied bands ${\cal A}_\mu = \sum_{n;E_n({\bf k})<0} \left<\Phi_n({\bf k})|\partial_{k_\mu}
  \Phi_n({\bf k})\right>$ ($\mu = x, y$). Integrating the Berry curvature $\nabla\times{\cal A}$ shown in Figure
  \ref{fig:c2}(a) over the Brillouin zone \cite{bib:hatsugai}, we find that $c = -1$ when the Zeeman field satisfies
  $\left[(\mu+4t_0)^2+\Delta^2\right]^{1/2} < V_z < \left(\mu^2+\Delta^2\right)^{1/2}$ with $\mu \leq -2t_0$, which is
  attributed to the topologically nontrivial energy gap at $(k_x,k_y) = (0,0)$ while those at $(0,-\pi/a)$, $(-\pi/a,0)$
  and $(-\pi/a,-\pi/a)$ remain trivial \cite{bib:alicea,bib:lutchyn,bib:sato_chern}. The nonzero Chern number indicates
  that the system is topological and thus may support MFs. The above result is consistent with the approximation
  treatment around $\Gamma$ point \cite{bib:sato_chern}.

\textbf{Core Majorana fermions.}
  We study a finite sample with two separated holes in the SM and two vortices pinned right beneath them (see top panel
  of Figure \ref{fig:c2}(b)). The typical size of a square sample is 300nm$\times$300nm, which is divided into $200\times 200$
  square grids, corresponding to Hamiltonian matrix of dimension $10^5\times 10^5$ in (\ref{eq:BdG}).
   By solving the BdG equation for this case, we obtain the energy spectra of excitations
  both at the vortex cores and the sample edge. Two zero-energy states are found at the holes, whereas no such state at
  the edge, as shown in Figure \ref{fig:c2}(c). We examine the four spinor components of the zero-energy states, and find
  $u_\uparrow = v_\uparrow^*$ and $u_\downarrow = v_\downarrow^*$ (displayed explicitly in Figure \ref{fig:c2}(d) only
  for the right hole), which results in $\gamma^\dagger = \gamma$, indicating that the two zero-energy states are
  Majorana states.

  It should be noticed that the excitation energy gap at vortices is about four times larger than that at the edge (see
  \ref{fig:c2}(c)), which makes the core MFs more stable than edge MFs \cite{bib:liang}. On the other
  hand, because the minigap associated with Andreev bound states at superconducting vortex core proximity-induced in SM is roughly $\Delta^2/(\Delta^2+\mu^2)^{1/2} \sim\Delta$ with a small Fermi energy $\mu\sim\Delta$ \cite{bib:minigap}, the influence from Andreev bound states to the core MFs can be neglected. It is in contrast to the case of SC exposed to vacuum where $\mu\gg\Delta$ and thus the minigap is small in order of $\Delta^2/\mu$.

  \begin{figure*}[ht]
    \centering
    \includegraphics[width=\textwidth]{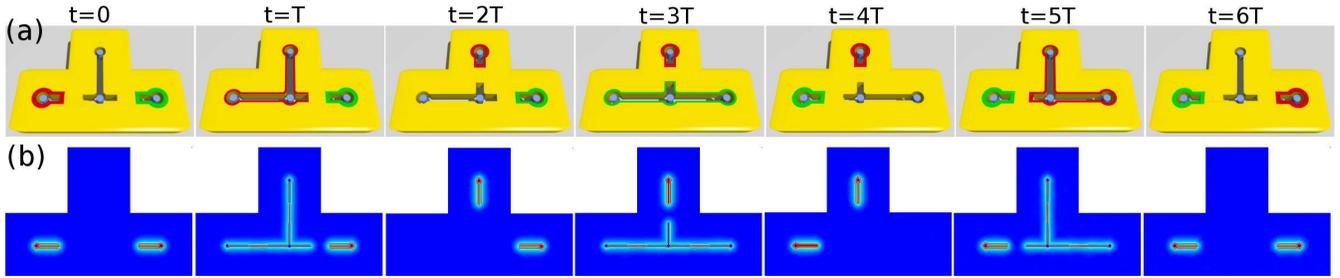}
      \caption{(a) Schematic diagram for braiding the two MFs at the left and right holes, colored by red and green
      respectively. (b) Distributions of the wave-functions of MFs obtained by solving TDBdG at several typical time steps.}
    \label{fig:braid}
  \end{figure*}

  Next, we impose a point-like gate voltage on the region between the two holes to prohibit direct hopping of electrons
  by lifting the on-site energy there as in the bottom panel of Figure \ref{fig:c2}(b) (see also Figure \ref{fig:setup}), which merges
  effectively the two isolated holes into a unified one. Solving the BdG equation for this case, there is no zero-energy
  quasi-particle excitation state, since the combined hole includes two vortices \cite{bib:it}.

  Based on the above result, we can design a way to liberate and transport a MF from one vortex to another (see
  the first three columns in Figure \ref{fig:braid}). Initially, the top and middle holes are connected together while
  the leftmost one is isolated and hosts a MF. We then combine these three holes by applying gate voltages on
  the region between the left and middle ones, which causes the MF to spread itself over the unified hole including
  three vortices ($t = T$). Finally, the MF is moved totally to the top by disconnecting the top hole from
  others ($t = 2T$). It is noticed that the collapsing of MF wave-function on the top hole is a topological property,
  and is impossible for electrons. Solving BdG equation numerically, we confirm that the
  gap between the zero-energy MF states and the excitations remains finite during the whole process, which
  guarantees the topological protection.

  Being able to transport one MF from one hole to another, we proceed to exchange positions of two
  core MFs in the system shown in Figure \ref{fig:braid} (see also Figure \ref{fig:setup}). Following the above transporting
  procedures, we further move the green MF from the right hole to the left one during $t = 2T\sim 4T$. At last, the
  red MF stored temporarily at the top hole is transported to the right one in the period $t = 4T\sim 6T$. After the sequence
  of switching processes, the system comes back to the original state with the red and green MFs exchanged.

  In order to investigate the impact of the position exchange to the MF state, we monitor the time evolution of MF wave functions
  $\left|\Psi(t)\right>$ by solving the TDBdG equation numerically
  \begin{equation}
    i\hbar\frac{d}{d t}\left|\Psi(t)\right> = H(t) \left|\Psi(t)\right>,
  \end{equation}
  where $H(t)$ is given in Figure \ref{eq:BdG} and depends on time via the hopping rates $t_{\bdi\bdj}$ and
  $t_\alpha^\bdi$ in the regions between holes, which are tuned adiabatically by the local gate voltages
  \cite{bib:liang}.

  Even with the powerful computation resources available in these days, it is still hopeless to tackle this problem by
  directly diagonalizing the Hamiltonian $H(t)$ of dimension $10^6\times 10^6$ for each time instant.  Fortunately, it
  has been revealed that when the exponential operator is expanded by the Chebyshev polynomial
  \begin{equation}
    \exp[-i H(t)\delta t/\hbar] = \sum_n c_n(\delta t) T_n(H),
    \label{eq:cheby}
  \end{equation}
  the coefficient $c_n(\delta t)$ decreases with $n$ exponentially fast for small $\delta t$\cite{bib:loh}.  Therefore,
  only several leading terms $c_n$ are necessary for a sufficiently accurate estimate of $\exp[-i H(t)\delta
  t/\hbar]$. Moreover, the Chebyshev polynomials can be constructed based on the recursive relation $T_n(H) =
  2HT_{n-1}(H) - T_{n-2}(H)$, which reduces the computation time further. In this way, the time-dependent wave-function
  of the MFs can be obtained efficiently in an iterative way $\left|\Psi(t+\delta t)\right> \simeq \exp[-i H(t)\delta
  t/\hbar]\left|\Psi(t)\right>$.

\begin{figure}[htbp]
    \centering
    \includegraphics[width=\linewidth]{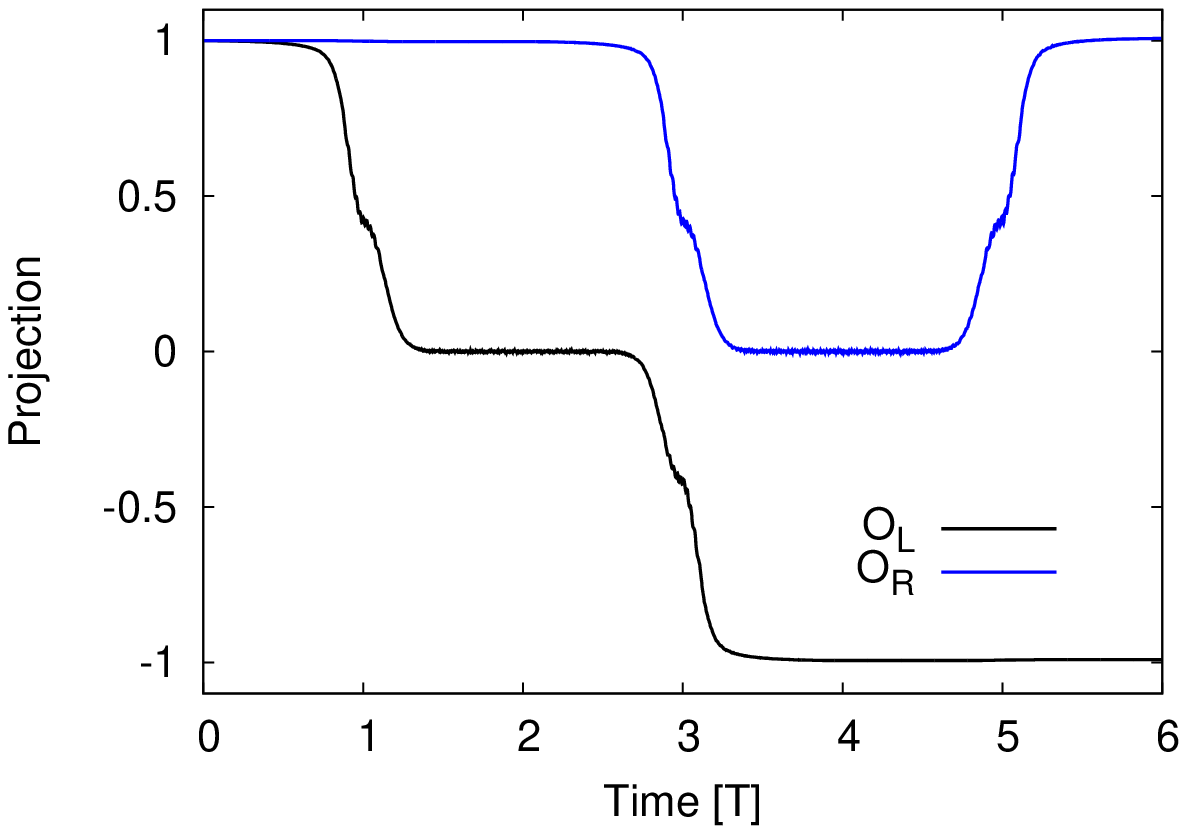}
      \caption{Projections of the MF wave-function $\left|\Psi(t)\right>$ obtained by TDBdG
      onto the initial states \(O_L =\left<\Psi_L(0)|\Psi(t)\right>\) and \(O_R = \left<\Psi_R(0)|\Psi(t)\right>\).}
      \label{fig:proj}
  \end{figure}

  The wave function $\left|\Psi(t=0)\right>$ is the superposition of the zero-energy states $\left|\Psi_L(0)\right>$
  and $\left|\Psi_R(0)\right>$ at left and right holes respectively as shown in Figure \ref{fig:braid}(a). We evaluate
  the projections of $\left|\Psi(t)\right>$ onto the initial states
  $O_L = \left<\Psi_L(0)|\Psi(t)\right>$ and $O_R = \left<\Psi_R(0)|\Psi(t)\right>$ and
  display them in Figure \ref{fig:proj} for $0 \le t \le 6T$. The two MFs pick up opposite signs after exchanging
  their positions, which can be summarized by
   \begin{equation}
   \gamma_L \rightarrow  \gamma_R, \quad \gamma_R \rightarrow -\gamma_L.
    \label{eq:na}
   \end{equation}
  Since (\ref{eq:na}) can be written as $\gamma_{L/R} \rightarrow U^\dagger\gamma_{L/R}U$ with the unitary matrix $U
  = \exp(\pi\gamma_R\gamma_L/4)$, the braiding of MFs satisfies non-Abelian statistics
  \cite{bib:sau_nano,bib:heck,bib:liang,bib:alicea_wire}.

  The braiding rule (\ref{eq:na}) can also be understood by the three pairs of opposite motions of MFs across links
  between holes during the whole process displayed in Figure \ref{fig:braid}, since each pair of opposite motions
  contributes -1 and $(-1)^3=-1$, which indicating that either $\gamma_L$ or $\gamma_R$ has to change sign
  \cite{bib:liang}.

\begin{figure}[t]
    \centering
    \includegraphics[width=.85\linewidth]{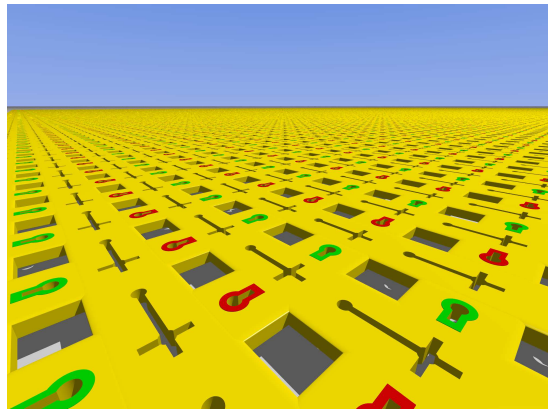}
    \caption{Device array for the present scheme for manipulation of MFs.}
    \label{fig:fig5}
  \end{figure}

  The above scheme for the braiding of MFs shows good scalability.  For a device in Figure \ref{fig:fig5} constructed by
  the unit displayed in Figure \ref{fig:setup}, one can exchange any two MFs by a sequence of switchings of point-like
  gate voltages. It is straightforward to figure out that any MF-pair exchanging accompanies odd number of opposite
  motions of MFs across links between holes, which guarantees the non-Abelian statistics \cite{bib:liang}. Therefore,
  the present scheme has the potential to be used for large scale manipulation of topological qubits.

\textbf{Conclusion.}
  We have shown that the Majorana fermions hosted by vortex cores in topological superconductors can be
  liberated from pinned vortices, transported and braided over the prepared holes, taking advantages of the
  heterostructure of \textit{s}-wave superconductor and spin-orbital coupling semiconductor. By solving the time-dependent
  Bogoliubov-de Gennes equation numerically, we monitor the time evolutions of Majorana fermion wave-functions and
  demonstrate the non-Abelian statistics of adiabatical braidings of Majorana fermions. The present scheme only
  requires local applications of gate voltages, and minimizes possible disturbances to the Majorana fermions, which
  might be a challenging issue in other proposals based on end Majorana fermions in one-dimensional superconductors
  where gate voltages are necessary along the whole system. As compared with the edge Majorana fermions in
  two-dimensional topological superconductors, the core Majorana fermions are protected by a larger energy gap, and
  thus reduce the limitation on operating temperatures. Therefore, the present scheme provides a more feasible way for
  manipulating Majorana fermions in large scale.

  \begin{acknowledgement}
  X.H. thanks Y. Kato for useful discussions. This work was supported by WPI initiative on Materials
  Nanoarchitectonics, MEXT of Japan.
  \end{acknowledgement}


\begin{thebibliography}{99}
  \bibitem{bib:em}
    Majorana, E. \textit{Nuovo Cimento} {\bf 1937}, 5, 171.
  \bibitem{bib:kitaev2}
    Kitaev, A. Y. \textit{Ann. Phys.} {\bf 2003}, 303, 2.
  \bibitem{bib:cn_rmp}
    Nayak, C.; Simon, S. H.; Stern, A.; Freedman, M.; Das Sarma, S. \textit{Rev. Mod. Phys.} {\bf 2008}, 80, 1083.
  \bibitem{bib:wilczek}
    Wilczek, F. \textit{Nature Phys.} {\bf 2009}, 5, 614.
  \bibitem{bib:moore}
    Moore, G.; Read, N. \textit{Nucl. Phys. B} {\bf 1991}, 360, 362.
  \bibitem{bib:read}
    Read, N.; Green, D. \textit{Phys. Rev. B} {\bf 2000}, 61, 10267.
  \bibitem{bib:kitaev1}
    Kitaev, A. Y. \textit{Phys. -Usp.} {\bf 2001}, 44, 131.
  \bibitem{bib:tewari}
    Tewari, S.; Das Sarma, S.; Nayak, C.; Zhang, C.-W.; Zoller, P. \textit{Phys. Rev. Lett.} {\bf 2007}, 98, 010506.
  \bibitem{bib:sato}
    Sato, M.; Takahashi, Y.; Fujimoto, S. \textit{Phys. Rev. Lett.} {\bf 2009}, 103, 020401.
  \bibitem{bib:fu_ti}
    Fu, L.; Kane, C. L. \textit{Phys. Rev. Lett.} {\bf 2008}, 100, 096407.
  \bibitem{bib:sau_dev}
    Sau, J. D.; Lutchyn, R. M.; Tewari, S.; Das Sarma, S. \textit{Phys. Rev. Lett.} {\bf 2010}, 104, 040502.
  \bibitem{bib:alicea}
    Alicea, J. \textit{Phys. Rev. B} {\bf 2010}, 81, 125318.
  \bibitem{bib:linder}
    Linder, J.; Tanaka, Y.; Yokoyama, T.; Sudb\o, A.; Nagaosa, N. \textit{Phys. Rev. Lett.} {\bf 2010}, 104, 067001.
  \bibitem{bib:lutchyn}
    Lutchyn, R. M.; Sau, J. D.; Das Sarma, S. \textit{Phys. Rev. Lett.} {\bf 2010}, 105, 077001.
  \bibitem{bib:oreg}
    Oreg, Y.; Rafael, G.; Oppen, F. von \textit{Phys. Rev. Lett.} {\bf 2010}, 105, 177002.
  \bibitem{bib:hassler}
    Hassler, F.; Akhmerov, A. R.; Hou, C.-Y.; Beenakker, C. W. J. \textit{New J. Phys.} {\bf 2010}, 12, 125002.
  \bibitem{bib:potter}
    Potter, A. C.; Lee, P. A. \textit{Phys. Rev. Lett.} {\bf 2010}, 105, 227003.
  \bibitem{bib:sau_nano}
    Sau, J. D.; Clarke, D. J.; Tewari, S. \textit{Phys. Rev. B} {\bf 2011}, 84, 094505.
  \bibitem{bib:heck}
    Heck, B. van; Akhmerov, A. R.; Hassler, F.; Burrello, M.; Beenakker, C. W. J. \textit{New J. Phys.} {\bf 2012}, 14, 035019.
  \bibitem{bib:mourik}
    Mourik, V.; Zuo, K.; Frolov, S. M.; Plissard, S. R.; Bakkers, E. P. A. M.; Kouwenhoven, L. P. \textit{Science} {\bf
    2012}, 336, 1003.
  \bibitem{bib:ivanov}
    Ivanov, D. A. \textit{Phys. Rev. Lett.} {\bf 2001}, 86, 268.
  \bibitem{bib:liang}
    Liang, Q.-F.; Wang Z.; Hu, X. \textit{ Europhys. Lett.} {\bf 2012}, 99, 50004.
  \bibitem{bib:TKNN}
    Thouless, D. J.; Kohmoto, M.; Nightingale, M. P.; Nijs, M. den \textit{Phys. Rev. Lett.} {\bf  1982}, 49, 405.
  \bibitem{bib:hatsugai}
    Fukui, T.; Hatsugai, Y.; Suzuki, H. \textit{J. Phys. Soc. Jpn.} {\bf 2005}, 74, 1674.
  \bibitem{bib:sato_chern}
    Sato, M.; Takahashi, Y.; Fujimoto, S. \textit{Phys. Rev. B} {\bf 2010}, 82, 134521.
  \bibitem{bib:minigap}
    Sau, J. D.; Lutchyn, R. M.; Tewari, S.; Sarma, S. Das \textit{Phys. Rev. B} {\bf 2010}, 82, 094522.
  \bibitem{bib:it}
    Tewari, S.; Das Sarma, S.; Lee, D.-H. \textit{Phys. Rev. Lett.} {\bf 2007}, 99, 037001.
  \bibitem{bib:loh}
    Loh, Y. L.; Taraskin, S. N.; Elliott, S. R. \textit{Phys. Rev. Lett.} {\bf 2000}, 84, 2290.
  \bibitem{bib:alicea_wire}
    Alicea, J.; Oreg, Y.; Refael, G.; Oppen, F. von; Fisher, M. P. A. \textit{Nature Phys.} {\bf 2011}, 7, 412.
\end{thebibliography}
\end{document}